\documentclass[aps,onecolumn,10pt]{revtex4}
\usepackage{amsmath}
\usepackage{amssymb}
\usepackage{graphicx}
\usepackage{hyperref}
\numberwithin{equation}{section}

\hypersetup{colorlinks=true,linkcolor=blue,citecolor=red} 
\begin{document}

\title{External Field Effect in Gravity}

\author{Philip D. Mannheim${}^1$ and  John W. Moffat${}^2$}
\affiliation{${}^1$Department of Physics, University of Connecticut, Storrs, CT 06269, USA \\
${}^2$Perimeter Institute for Theoretical Physics, Waterloo, Ontario N2L 2Y5, Canada\\
philip.mannheim@uconn.edu,~jmoffat@perimeterinstitute.ca}

\date{March 25 2021}

\begin{abstract}
In both Newtonian gravity and Einstein gravity there is no force on a test particle located inside a spherical cavity cut out of a static, spherically symmetric mass distribution. Inside the cavity exterior matter is decoupled and there is no external field effect that could act on the test particle. However, for potentials other than the Newtonian potential or for geometries other than Ricci flat ones this is no longer the case, and there then is an external field effect. We explore this possibility in various alternate gravity scenarios, and suggest that such (Machian) external field effects can serve as a diagnostic for gravitational theory.
\end{abstract}

\maketitle

\noindent
Essay written for the Gravity Research Foundation 2021 Awards for Essays on Gravitation.

\section{Introduction}
\label{S1}

For a static, spherically symmetric source $\rho(r)$ in Newtonian gravity the potential is given by
\begin{eqnarray}
\phi(r)=\int d^3r^{\prime} \frac{\rho(r^{\prime})}{|\vec{r}-\vec{r}^{\prime}|}.
\label{1.1e}
\end{eqnarray}
Since $\rho(r^{\prime})$ has no dependence on angle, on doing the angular integration we obtain
\begin{eqnarray}
\phi(r)=\frac{2\pi}{ r}\int r^{\prime }dr^{\prime} \rho(r^{\prime})[|r+r^{\prime}|-|r-r^{\prime}|].
\label{1.2e}
\end{eqnarray}
If $\rho(r^{\prime})$ is only nonzero in a  band of range $R_1<r^{\prime}<R_2$, in the $r>R_2$ and $r<R_1$ regions the potential is given by
\begin{eqnarray}
\phi(r>R_2)=\frac{4\pi}{r}\int _{R_1}^{R_2}r^{\prime 2}dr^{\prime} \rho(r^{\prime}), \quad
\phi(r<R_1)=4\pi\int _{R_1}^{R_2}r^{\prime }dr^{\prime} \rho(r^{\prime}).
\label{1.3e}
\end{eqnarray}
 Thus as well as recover the standard $1/r$ behavior outside the band, we see that in the $r<R_1$ cavity the potential is constant and the force is zero.  Moreover, in $r<R_1$ the potential remains constant even if we let $R_2$ go to infinity. Thus no matter how complicated $\rho(r^{\prime})$ might be, and no matter how far it may extend, as long as it vanishes in $r^{\prime}<R_1$ a static spherically symmetric exterior does not affect the force on a test particle inside the cavity. 

While we see that there is no external field effect in Newtonian gravity for spherically-symmetrically-distributed sources, we should note that the Universe is not completely spherically symmetric. There are inhomogeneities in the otherwise maximally 3-symmetric isotropic Hubble flow due to cosmological fluctuations that produce the distribution of matter sources that we observe, and these matter sources have peculiar velocities with respect to the Hubble flow. With these sources also producing Newtonian $1/r^2$ forces, their contributions are suppressed at large distances from a given system, with the more nearby sources only producing tidal forces that fall off as $1/r^3$. With this caveat we shall ignore such contributions and consider any external field effect due to them to be negligible. 

To contrast, let us instead consider a potential that behaves as $|\vec{r}-\vec{r}^{\prime}|^{2\alpha -1}$ where $\alpha$ is a constant. Now we obtain
\begin{eqnarray}
\phi(r)=\frac{2\pi}{ r(2\alpha+1)}\int r^{\prime }dr^{\prime} \rho(r^{\prime})[|r+r^{\prime}|^{2\alpha+1}-|r-r^{\prime}|^{2\alpha+1}],
\label{1.4e}
\end{eqnarray}
\begin{eqnarray}
\phi(r<R_1)=\frac{2\pi}{ r (2\alpha+1)}\int_{R_1}^{R_2} r^{\prime }dr^{\prime} \rho(r^{\prime})[(r+r^{\prime})^{2\alpha+1}-(r^{\prime}-r)^{2\alpha+1}],
\label{1.5e}
\end{eqnarray}
\begin{eqnarray}
\phi(r>R_2)=\frac{2\pi}{ r (2\alpha+1)}\int_{R_1}^{R_2} r^{\prime }dr^{\prime} \rho(r^{\prime})[(r+r^{\prime})^{2\alpha+1}-(r-r^{\prime})^{2\alpha+1}].
\label{1.6e}
\end{eqnarray}
As we see, for any  $\alpha\neq 0$ this time there is a dependence on $r$ inside the cavity, and not only that, there is a departure from $1/r$ outside the mass distribution. As an illustrative example, consider $\alpha=1$ (a case we will actually return to below). Then we have
\begin{eqnarray}
\phi(r<R_1)=\frac{4\pi}{3}\int_{R_1}^{R_2} dr^{\prime} \rho(r^{\prime})\left( r^2r^{\prime}+3r^{\prime 3}\right),
\label{1.7e}
\end{eqnarray}
\begin{eqnarray}
\phi(r>R_2)=\frac{4\pi}{3 }\int_{R_1}^{R_2} dr^{\prime} \rho(r^{\prime})\left( 3rr^{\prime 2}+\frac{r^{\prime 4}}{r}\right),
\label{1.8e}
\end{eqnarray}
with $\phi(r<R_1)$ not being constant, and $\phi(r>R_2)$ not being just of the $1/r$ form. As we see, when there is an external field effect inside the cavity there is also a departure from $1/r$ outside the mass distribution.

It is also instructive to consider the situation from the perspective of  the Poisson equation. If we consider the Green's function equation
\begin{eqnarray}
[\nabla^{2}]^{1+\alpha}G(\vec{r}-\vec{r}^{\prime})=\delta^3(\vec{r}-\vec{r}^{\prime}),
\label{1.9e}
\end{eqnarray}
then on Fourier transforming we find that $\tilde{G}(k)\sim (k^2)^{-1-\alpha}$, so that $G(\vec{r}-\vec{r}^{\prime})\sim |\vec{r}-\vec{r}^{\prime}|^{2\alpha-1}$. Thus we see that the $|\vec{r}-\vec{r}^{\prime}|^{2\alpha-1}$ potential is also the Green's function for the $[\nabla^{2}]^{1+\alpha}$ derivative operator. Thus there will always be an external field effect for any Poisson equation with $\alpha \neq 0$, or even more generally for any Poisson equation more complicated than the standard second order one.

As regards covariant metric theories of gravity, we recall that Einstein specifically chose second-order gravitational equations of motion for the metric so that they would reduce to the standard second-order Poisson equation in the non-relativistic limit. As such, the Einstein equations provide a generalization of the standard second-order Poisson equation, putting it into a form that all accelerating observers could agree on. We can thus infer that in any covariant theory of gravity other than standard Einstein gravity itself there will be an external field effect of some sort.

To see that there is no external field effect in Einstein gravity itself we study standard gravity with a static, spherically symmetric source. In general, such a source should not actually be taken to be the perfect fluid source that is commonly used, since while the assumption of maximally 2-symmetric spherical symmetry about only one single point (taken to be the origin of coordinates) requires that $T^{\theta}_{\phantom{\theta}\theta}=T^{\phi}_{\phantom{\phi}\phi}$, it does not require that $T^{r}_{\phantom{r}r}$ be equal to $T^{\theta}_{\phantom{\theta}\theta}$ \cite{Mannheim2006,Mannheim2010}. Now a general rank two tensor would have ten independent components, and they are typically taken to be two scalar components, $\rho$ and $p$, a three-vector heat flow that obeys $U^{\mu}q_{\mu}=0$ where $U^{\mu}$ is a unit timelike vector, and a five component traceless and symmetric viscosity tensor that obeys $U^{\mu}\pi_{\mu\nu}=0$.  Constructing these components by incoherently averaging over all the modes of a scalar field propagating in a static, maximally 2-symmetric spherically symmetric geometry shows \cite{Mannheim2010} that only three components are present, $\rho$, $p$ and a remnant of the viscosity, which we label $q$. Thus in order to discuss any possible external field effect in Einstein gravity in as general a way as possible and not miss any effect that the $q$ term might give rise to, we take the metric and the nonzero components of the energy-momentum tensor of the matter source to be of the form 
\begin{eqnarray}
&&ds^2=B(r)dt^2-A(r)dr^2-r^2d\theta^2-r^2\sin^2\theta d\phi^2, 
\nonumber\\
&&T_{00}=\rho(r) B(r),\quad T_{rr}=p(r) A(r),\quad T_{\theta\theta}=q(r)r^2,\quad T_{\phi\phi}=q(r)r^2{\rm sin}^2\theta. 
\label{1.10e}
\end{eqnarray}                                 
Covariant conservation of the energy-momentum tensor requires that \cite{Mannheim2006}
\begin{equation}
\frac{dp}{dr}+\frac{(\rho +p)}{2B}\frac{dB}{dr}+\frac{2}{r}(p-q)=0,
\label{1.11e}
\end{equation}                                 
while the Einstein field equations yield \cite{Mannheim2006}
\begin{equation}
A^{-1}(r)=1-\frac{2 G}{r}\tilde{M}(r),
\label{1.12e}
\end{equation}                                 
\begin{equation}
\frac{1}{B}\frac{dB}{dr}=\frac{2G}{r(r-2G\tilde{M})}
\left(\tilde{M}+4\pi r^3p\right),
\label{1.13e}
\end{equation}                                 
where 
\begin{equation}
\tilde{M}(r)=4\pi\int_0^rdr r^2\rho(r).
\label{1.14e}
\end{equation}                                 
We note that the equations for $A(r)$ and $B(r)$ are the same as they would have been had we taken the source to be a perfect fluid with $q(r)=p(r)$, with the  $p-q$ difference only appearing in the conservation condition. 

If we now confine the source to the range $R_1<r<R_2$, we see that for $r<R_1$ both $A(r)$ and $B(r)$ are constant. Moreover, this remains true even if we allow $R_2$ to go all the way to infinity, and is not affected by however complicated the structure of $\rho(r)$, $p(r)$ and $q(r)$ might be as long as they are all zero in $r<R_1$. Thus in parallel with its Newtonian limit, in Einstein gravity the geometry inside a static, spherical cavity is flat. There is thus no external field effect in Einstein gravity. Moreover, it was noted by Weinberg \cite{Weinberg1972} that even if the source is not static but still spherically symmetric then inside a spherical cavity within the source the geometry is also flat. With Birkhoff's theorem showing that outside the source the geometry is the same as if the source were static, the same holds inside a cavity within the source. Thus  in Einstein gravity even for time dependent, spherical symmetric systems there is no external field effect.

In contrast, in another possible covariant metric-based theory of gravity, viz. the fourth-order conformal gravity theory  that we discuss in more detail below, we note that the gravitational equations of motion reduce without approximation to a fourth-order Poisson equation $\nabla^4B(r)=f(r)$ with a source function $f(r)$ that is given in (\ref{3.2e}) \cite{Mannheim1994,Mannheim2006}. The theory thus falls into the $\alpha=1$ case studied above, with an external field effect immediately being apparent in (\ref{1.7e}). This example thus exhibits our central theme, namely that departures from either the standard second-order Poisson equation or from the standard second-order Einstein equations will result in an external field effect. To establish the ubiquity of this claim, in this paper we shall study three possible alternative gravity theories, the MOND (Modified Newtonian Dynamics) theory introduced by Milgrom \cite{Milgrom1983a,Milgrom1983b,Milgrom1983c}, the conformal gravity theory studied by Mannheim and Kazanas \cite{Mannheim1989,Mannheim1994}, and the MOG (Modified Gravity) theory developed by Moffat \cite{Moffat2006}. We shall establish that in all three of them there is an explicit external field effect. And regardless of how large or small such effects might be, it is their very existence that is of significance for gravity theory. We discuss these theories in the chronological sequence in which they were developed.

\section{The Modified Newtonian Dynamics MOND Theory}
\label{S2}

As it started to become apparent that the standard Newton-Einstein theory failed to account for a variety of astrophysical data if the only matter sources considered were the known luminous ones, the consensus of the community was to introduce more sources than had been detected, sources initially referred to as missing mass, and subsequently when no additional luminous sources were detected, as dark, i.e., intrinsically non-luminous, matter.  Since in the Newton-Einstein theory such dark matter sources would have to be located wherever any missing mass problem might be detected,  they would not only leave the standard Newton-Einstein theory intact, they would not cause it to acquire any external field effect. Now gravity theory is based on equations that relate a gravity side to a matter side, and in principle it is just as valid to consider modifying the gravity side as the matter side. As noted above, any attempt in this direction would then lead to an external field effect, and thus provide a diagnostic for any such modifications.

\begin{figure}[htpb!]
  \centering
  \includegraphics[width=3.2in,height=2.0in]{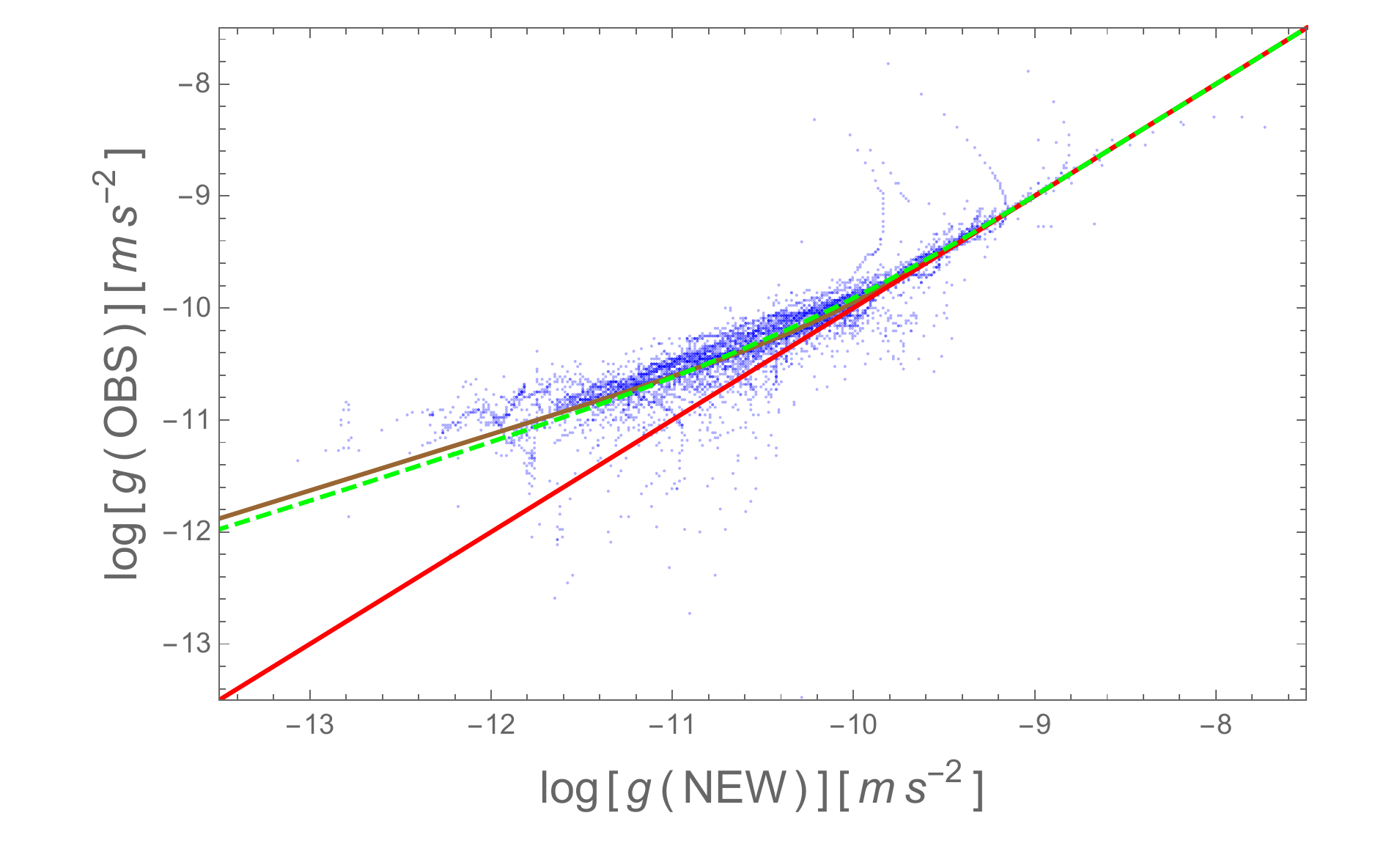}
  \caption{The $g(OBS)$ versus $g(NEW)$ plot and the $g(MLS)$  and $g(MOND)$ fits (dotted and full curves) to it. The solid diagonal is the line $g(OBS)=g(NEW)$.}
   \label{gobsgbar}
\end{figure}

Recognizing that something needed to be done if we did not want to resort to dark matter, Milgrom \cite{Milgrom1983a,Milgrom1983b,Milgrom1983c} proceeded empirically and extracted out certain generic features from the systematics of the galactic rotation curves (plots of the velocity of material in a given galaxy versus distance from its galactic center) that are observed in spiral galaxies. Using the quite small amount of data that was available at the time Milgrom noted a quite striking regularity, namely that the luminous Newtonian expectation for the rotational velocity $v$ failed whenever the acceleration $g(NEW)$ produced by the luminous Newtonian sources fell below a parameter  $a_0$ with the dimensions of acceleration. Over the years this result has held up as more and more spiral galaxy data have come on line. To exhibit this explicitly we note that in a recent study McGaugh, Lelli, and Schombert \cite{McGaugh2016} found a way to display all of the data points from a large number of galactic rotation curves on one single graph by plotting the measured $g(OBS)$ versus $g(NEW)$ for each data point (the so-called radial acceleration relation (RAR)). Following this approach and using a very big data sample (5791 total points from 207 galaxies of quite varying morphology) O'Brien, Chiarelli, and Mannheim generated \cite{O'Brien2018} a plot of the measured $g(OBS)$ versus $g(NEW)$ as shown in Fig. [\ref{gobsgbar}].
As can be seen from the figure, for the entire sample departures from $g(OBS)=g(NEW)$ set in when $g(NEW)$ drops below $10^{-10}$ m s$^{-2}$ or so.

In order to constrain the $g(NEW)<a_0$ region Milgrom appealed to two additional features in the data, namely that spiral velocities obey the empirical Tully-Fisher relation that $v^4/L$ is universal ($L$ being the galactic luminosity) and that rotation velocities should be asymptotically flat. However, rather than modify the gravity side of Newton's Law of gravitation Milgrom instead modified the inertial side (hence Modified Newtonian Dynamics) by setting 
\begin{equation}
 \mu\left(\frac{g(MOND)}{a_0}\right)g(MOND)=g(NEW),
\label{2.1e}
\end{equation}                                 
where $g(MOND)$ is the MOND prediction for the observed acceleration $g(OBS)$, $\phi$ is the ordinary Newtonian gravitational potential that obeys $\nabla^2\phi=-4\pi G\rho$ where $\rho$ is the luminous matter density, and $\mu(g(MOND)/a_0)$ is the modification. The choice
\begin{equation}
\mu(x)= \frac{x}{(1+x^2)^{1/2}}
\label{2.2e}
\end{equation}                                 
then fits all of the above requirements and has had much phenomenological success \cite{Sanders2002}; and when $\phi$ is integrated over the luminous material within the galaxies it leads to the solid curve shown in Fig. [\ref{gobsgbar}]. And with the mass to light ratios and adopted distances to galaxies used in \cite{O'Brien2018} $a_0$ was found to evaluate to $a_0=0.6\times 10^{-10}$ m s$^{-2}$. In their presentation the authors of \cite{McGaugh2016} provided an alternate fitting  formula
\begin{eqnarray}
 g(MLS)=\frac{g(NEW)}{[1-\exp(-(g(NEW)/g_0)^{1/2})]}, 
 \label{2.3e}
 \end{eqnarray}
and from the dotted curve shown in Fig. [\ref{gobsgbar}] the authors of \cite{O'Brien2018} extracted a value $g_0=0.6\times 10^{-10}$ m s$^{-2}$. Both sets of fits thus establish the existence of a universal acceleration scale of order $10^{-10}$ m s$^{-2}$.

For our purposes here we note that in \cite{Bekenstein1984} it was shown that the form given in (\ref{2.1e}) can be associated with a Poisson equation for the gravitational potential of the non-linear form 
\begin{equation}
 \vec{\nabla}\cdot \left[\mu\left(\frac{|\vec{\nabla}\phi |}{a_0}\right)\vec{\nabla}\phi\right]=4\pi G\rho.
\label{2.4e}
\end{equation}                                 
Since this equation is modified from the standard second-order Poisson equation, then just as we noted above, there will be an external field effect. And that there would be such an effect was already noted in \cite{Bekenstein1984}. In \cite{Bekenstein1984} a first attempt at a relativistic theory of MOND was made. A more evolved form was presented in \cite{Bekenstein2004}, and with it containing scalar, vector and tensor fields it is known as TeVeS. Interestingly, the MOG theory that we discuss in Sec. \ref{S4} below is also based on scalar, vector and tensor fields. Since TeVeS goes beyond pure tensor field Einstein gravity it too would lead to an external field effect.

As we had noted above, MOND fits such as the one given in Fig. [\ref{gobsgbar}] only used the luminous sources within each galaxy, with no external field contribution being needed. However, a very recent study of the Fig. [\ref{gobsgbar}] plot has found  \cite{Chae2020} that the fit can be statistically improved if an external field effect is in fact included. In this fit the authors of \cite{Chae2020} did not use the standard MOND formula given in (\ref{2.2e}) but instead used a sometimes considered alternate one of the form $\mu(x)=x/(1+x)$. With this choice the external field effect acceleration $g(EFE)$ is predicted to be of the form \cite{Chae2020}
\begin{equation}
g(EFE)=\nu\left(\frac{g(NEW)}{a_0}\right)g(NEW),\quad v(y)=\frac{1}{2}-\frac{A_e}{y}+\left[\frac{1+e}{y}+\left(\frac{1}{2}-\frac{A_e}{y}\right)^2\right]^{1/2},\quad
A_e=\frac{e(2+e)}{2(1+e)}, 
\label{2.5e}
\end{equation}
where $e=g(EXT)/a_0$ is a measure of the external field contribution as expressed in terms of an effective external acceleration $g(EXT)$. On fitting Fig. [\ref{gobsgbar}] with (\ref{2.5e}) the authors of \cite{Chae2020} found that $e=0.033$ is statistically preferred over $e=0$, with the largest departures from $e=0$ being found in the low $g(NEW)$ region.  While this study thus does in principle provide some support for an external field effect of order 3 per cent or so, we turn now to a theory (conformal gravity) in which an external field effect is not only built in, the theory provides support for a significant external field effect over the entire $g(OBS)>g(NEW)$ region exhibited in Fig. [\ref{gobsgbar}].

\section{Conformal Gravity}
\label{S3}

Conformal gravity is based on the action $I_{\rm W}=-\alpha_g\int d^4x(-g)^{1/2}C_{\lambda\mu\nu\kappa}
C^{\lambda\mu\nu\kappa}$, where $C_{\lambda\mu\nu\kappa}$ is the Weyl tensor and $\alpha_g$ is a dimensionless  gravitational coupling constant. As constructed, this action is the unique polynomial action in four spacetime dimensions that is invariant under local conformal scalings of the metric of the form $g_{\mu\nu}(x)\rightarrow e^{2\alpha(x)}g_{\mu\nu}(x)$, where $\alpha(x)$ is an arbitrary function of $x$. For this action the  conformal gravity gravitational equations of motion can be written in the compact form \cite{Mannheim2006}
\begin{equation}                                                                               
4\alpha_g[2\nabla_{\kappa}\nabla_{\lambda}C^{\mu\lambda\nu\kappa}-R_{\lambda \kappa}C^{\mu\lambda\nu\kappa}]=T^{\mu\nu}.
\label{3.1e}
\end{equation}      
In the static, spherically symmetric situation (\ref{3.1e}) can be solved exactly \cite{Mannheim1989,Mannheim1994}, where for such geometries  the imposition of conformal invariance on the metric in (\ref{1.10e}) allows us to set $A=1/B$ as a kinematic condition.  With $A=1/B$ the gravitational equations of motion associated with the metric given in (\ref{1.10e}) reduce without any approximation at all to \cite{Mannheim1994} the remarkably compact
\begin{equation}                                                                               
B^{\prime\prime\prime\prime}+\frac{4}{r}B^{\prime\prime\prime}= \nabla^4 B(r) = \frac{3}{4\alpha_g B(r)}\left(T^0_{{\phantom 0} 0}
-T^r_{{\phantom r} r}\right) =f(r),
\label{3.2e}
\end{equation}      
with  (\ref{3.2e}) serving to define $f(r)$. As such, (\ref{3.2e}) is an exact all-order relation in classical conformal gravity.

The solution to (\ref{3.2e}) can be determined in closed form and is given by  \cite{Mannheim1994} 
\begin{eqnarray}
B(r)&=&1-\frac{r}{2}\int_0^r
dr^{\prime}r^{\prime 2}f(r^{\prime})
-\frac{1}{6r}\int_0^r
dr^{\prime}r^{\prime 4}f(r^{\prime})
-\frac{1}{2}\int_r^{\infty}
dr^{\prime}r^{\prime 3}f(r^{\prime})
-\frac{r^2}{6}\int_r^{\infty}
dr^{\prime}r^{\prime }f(r^{\prime}),
\label{3.3e}
\end{eqnarray}                                 
where the $B(r)=1$ term satisfies $\nabla^4B(r)=0$.
If the source $f(r)$ is only nonzero in the ranges $(0,R_0)$, $(R_1,\infty)$,  on dropping a constant term the metric in the intermediate range $R_0<r<R_1$ takes the form
\begin{eqnarray}
B(R_0<r<R_1)=1-\frac{2\beta^*}{r}+\gamma^*r-\kappa r^2,
\label{3.4e}
\end{eqnarray}                                 
where
\begin{eqnarray}
\beta^*= \frac{1}{12}\int_0^{R_0}
dr^{\prime}r^{\prime 4}f(r^{\prime}),\qquad
\gamma^*=-\frac{1}{2}\int_0^{R_0}
dr^{\prime}r^{\prime 2}f(r^{\prime}),\quad \kappa=\frac{1}{6}\int_{R_1}^{\infty}
dr^{\prime}r^{\prime }f(r^{\prime}).
\label{3.5e}
\end{eqnarray}                                 
The $\beta^*$ and $\gamma^*$ terms refer to a solar mass star and were introduced in \cite{Mannheim1989}. Since sources are putting out potentials that grow with distance (viz. the $\gamma^* r$ type term), one cannot ignore the contribution of material outside of any system of interest. There are two forms of such external contributions, one due to the background cosmological Hubble flow and the other due to the inhomogeneities in it. The contribution of the inhomogeneities is accounted for by the $\kappa$ term \cite{Mannheim2011}. While the contribution of the cosmological background is not apparent in (\ref{3.3e}), in \cite{Mannheim1989,Mannheim1997} it was shown that when transformed by a coordinate transformation to any rest frame coordinate system with metric (\ref{1.10e}) a conformally transformed  global Robertson-Walker metric takes the form of a universal linear potential term $\gamma_0r$ contribution to $B(r)$, where $\gamma_0$ is fixed by the spatial curvature $k$ of the Universe according to $\gamma_0=(-4k)^{1/2}$. Support for a thereby required negative value for $k$ in conformal gravity is provided by a fully acceptable $k<0$, non-fine-tuned, non-dark-matter fit to the cosmological accelerating universe data \cite{Mannheim2006}. In the fit the contribution of the cosmological constant is under control precisely because of the underlying conformal symmetry. In fact, the original motivation \cite{Mannheim1990} for studying conformal gravity was because of this aspect of the theory, not the dark matter problem. That the theory is  then, as we now describe, able to handle the dark matter problem is an encouraging figure of merit for it.

With the inclusion of the $\gamma_0r$  term the intermediate region metric takes the form 
\begin{eqnarray}
B(R_0<r<R_1)&=&1-\frac{2\beta^*}{r}+\gamma^* r +\gamma_0 r-\kappa r^2.
\label{3.6f}
\end{eqnarray}
When integrated over the distribution of $N^*$ luminous stars and gas within each galaxy, 
the use of (\ref{3.6f}) leads to very good fitting to the rotation curves of 138 galaxies \cite{Mannheim2011,Mannheim2012,O'Brien2012} with fixed, universal (i.e., galaxy-independent) parameters
\begin{eqnarray}
\beta^*&=&1.48\times 10^5 {\rm cm},\quad \gamma^*=5.42\times 10^{-41} {\rm cm}^{-1},
\nonumber\\
\gamma_0&=&3.06\times
10^{-30} {\rm cm}^{-1},\quad \kappa = 9.54\times 10^{-54} {\rm cm}^{-2},
\label{3.7f}
\end{eqnarray} 
and with there being no need to introduce any dark matter. Since current dark matter fits require two free parameters per galactic halo, the galaxy-dependent 276 free dark matter halo parameters that are needed for the 138 galaxy sample are replaced by just the three universal parameters: $\gamma^*$, $\gamma_0$ and $\kappa$. (The luminous Newtonian $N^*\beta^*$ contribution associated with (\ref{3.6f}) is common to both dark matter and conformal gravity fits and is included in both cases.) With $\gamma_0$ being fitted to be of order the inverse of the Hubble radius and with the fitted $\kappa$ being of order a typical cluster of galaxies scale, the values for $\gamma_0$ and $\kappa$ that are obtained show that they are indeed of the cosmological scales associated with the homogeneous Hubble flow and the inhomogeneities in it. Since these terms come from the luminous material outside of any given galaxy they are bona fide external field effects that are, as already made apparent in (\ref{1.7e}) above, part and parcel of the fourth-order derivative conformal gravity theory. From the perspective of conformal gravity the missing mass shortfall in galaxies is explained by the rest of the visible mass in the universe. The missing mass is thus not missing at all, it is the rest of the visible universe and it has been hiding in plain sight all along. 

\begin{figure}[htpb!]
  \centering
    \includegraphics[width=3.2in,height=2.0in]{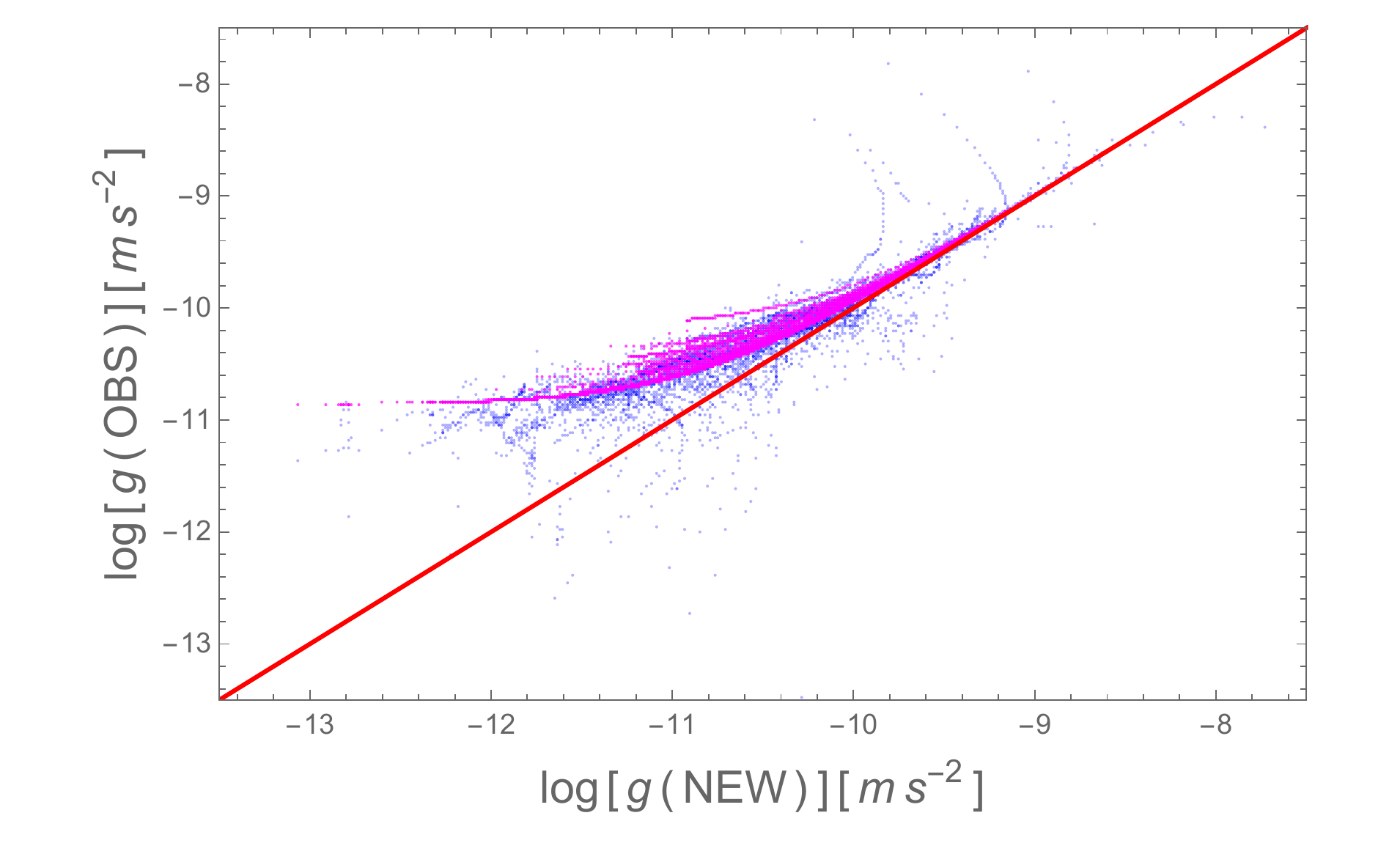}
  \caption{Conformal gravity prediction for $g(OBS)$ versus $g(NEW)$. The solid lines other than the diagonal are the conformal gravity expectations. Unlike in the single line $g(MLS)$ and $g(MOND)$ fits the conformal gravity fits form a band that conforms to the observed spread in the data points.}
   \label{cgoverlayall}
\end{figure}
In Fig. [\ref{cgoverlayall}] we present the  $g(OBS)$ versus $g(NEW)$ plot in the conformal gravity theory. As we see, despite the fact that there are so few parameters in (\ref{3.6f}), conformal gravity captures the data. The external $\gamma_0$ and $\kappa$ terms make significant contributions to the data over the entire region where $g(OBS)$ differs from $g(NEW)$. There is thus an  effect on local systems coming from global physics in every part of the $g(OBS)>g(NEW)$ region of the plot. and it can be thought of as a realization of Mach's principle.

\section{The Modified Gravity MOG Theory}
\label{S4}

The MOG theory is a scalar, vector, tensor theory involving two scalar fields $G$ and $\mu$, a vector field $\phi_{\mu}$ and the tensor metric $g_{\mu\nu}$. Its action is of the form \cite{Moffat2006}:
\begin{align}
S=S_G+S_\phi+S_S+S_M,
\end{align}
where $S_M$ is the matter action and
\begin{align}
  S_G&=\frac{1}{16\pi}\int d^4x(-g)^{1/2}\left[\frac{1}{G}(R+2\Lambda)\right],\\
  S_\phi&=\int d^4x(-g)^{1/2}\left[-\frac{1}{4}B^{\mu\nu}B_{\mu\nu}+\frac{1}{2}\mu^2\phi^\mu\phi_\mu\right],\\
  S_S&=\int
       d^4x(-g)^{1/2}\left[\frac{1}{G^3}\left(\frac{1}{2}g^{\mu\nu}\partial_\mu
       G\partial_\nu G-V_G\right)+\frac{1}{\mu^2G}\left(\frac{1}{2}g^{\alpha\beta}\partial_\alpha\mu\partial_\beta\mu-V_\mu\right)\right].\label{Eq:SS}
\end{align}
Here $B_{\mu\nu}=\partial_\mu\phi_\nu-\partial_\nu\phi_\mu$, $V_G$
and $V_\mu$ are potentials, $R$ is the Ricci scalar, and $\Lambda$ is a constant. The matter energy-momentum tensor is obtained by varying the matter
action $S_M$ with respect to the metric according to $T^{\mu\nu}_M=-2(-g)^{-1/2}\delta S_M/\delta g_{\mu\nu}$, while varying $S_\phi+S_S$ with respect to the metric yields
\begin{align}
  T^{\mu\nu}_{\rm MOG}=-2(-g)^{-1/2}[\delta S_\phi/\delta
  g_{\mu\nu}+\delta S_S/\delta g_{\mu\nu}],
\end{align}
and a thus total energy-momentum tensor $ T^{\mu\nu}=T^{\mu\nu}_{\rm M}+T^{\mu\nu}_{\rm MOG}$.
The MOG field equations are given by
\begin{align}
  R_{\mu\nu}-\frac{1}{2} g_{\mu\nu} R-\Lambda g_{\mu\nu}+Q_{\mu\nu}=8\pi GT_{\mu\nu}\,,
  \label{eq:MOGE}
\end{align}
\begin{equation}
  \frac{1}{(-g)^{1/2}}\partial_\mu\biggl((-g)^{1/2}B^{\mu\nu}\biggr)+\mu^2\phi^\nu=-J^\nu,
  \label{eq:phi}
\end{equation}
\begin{equation}
  \partial_\sigma B_{\mu\nu}+\partial_\mu B_{\nu\sigma}+\partial_\nu B_{\sigma\mu}=0,
\end{equation}
where
\begin{align}
  Q_{\mu\nu}=\frac{2}{G^2}(\partial^\alpha G \partial_\alpha G\,g_{\mu\nu}
  - \partial_\mu G\partial_\nu G) - \frac{1}{G}(g_{\mu\nu}\nabla_{\alpha}\nabla^{\alpha} G
  - \nabla_\mu\nabla_\nu G),\quad J_\mu=-(-g)^{-1/2}\frac{\delta S_M}{\delta \phi^\mu}.
  \label{eq:Q}
\end{align}
Combining the Bianchi identity $\nabla_\nu[R^{\mu\nu}-\frac{1}{2} g^{\mu\nu} R]=0$ with the
field equation (\ref{eq:MOGE}) yields the conservation law
\begin{align}
  \nabla_\nu T^{\mu\nu}+\frac{1}{G}\nabla_\nu G\,T^{\mu\nu} -
  \frac{1}{8\pi G}\nabla_\nu Q^{\mu\nu}=0 \,.
  \label{eq:Conservation}
\end{align}

In MOG all baryonic matter sources possess, in
proportion to their mass $M$, positive gravitational charge:
$Q_g=\kappa\,M$.  This charge serves as the source of the vector field
$\phi^\mu$.  Moreover, $\kappa=(G-G_N)^{1/2}=(\alpha\,G_N)^{1/2}$, where
$G_N$ is Newton's gravitational constant and
$\alpha=(G-G_N)/G_N\ge 0$.  

The equation of motion for a massive test particle in MOG has a
covariant form  \cite{Moffat2006} that is familiar from electromagnetism:
\begin{equation}
\label{eqMotion}
m\biggl(\frac{du^\mu}{ds}+\Gamma^{\mu}_{\alpha\beta}u^\alpha
u^\beta\biggr)= q_g{B^\mu}_\nu u^\nu,
\end{equation}
where $u^\mu=dx^\mu/ds$, $s$ the proper time along the particle
trajectory, and $m$ and $q_g$ denote the test particle mass $m$ and
gravitational charge $q_g=(\alpha G_N)^{1/2}m$.  We note
that for $q_g/m=(\alpha G_N)^{1/2}$ the equation of motion for a
massive test particle (\ref{eqMotion}) satisfies the (weak)
equivalence principle, with massive test particles being in mass-independent free fall in a homogeneous gravitational field even though they do not follow geodesics.
Photons have $m_\gamma=0$ and satisfy the null geodesic equation $k^\mu\nabla_\mu k^\nu=0$ where $k^\mu$ is the 4-momentum vector of photons and $k^2=g^{\mu\nu}k_\mu k_\nu=0$. Gravitational waves follow the same null geodesics as photons~\cite{GreenMoffatToth2018}.

Under the above conditions the modified Newtonian
acceleration law for a point particle can be written
as \cite{Moffat2006}:
\begin{equation}
\label{MOGaccelerationlaw}
a_{\rm MOG}(r)=-\frac{G_NM}{r^2}[1+\alpha-\alpha e^{-\mu r}(1+\mu r)].
\end{equation}
This reduces to Newton's gravitational acceleration in the limit
$\mu r\ll 1$.

The modified MOG acceleration equation (\ref{MOGaccelerationlaw}) fits galactic  rotation curves and galaxy cluster dynamics~\cite{MoffatRahvar2013,MoffatRahvar2014,Brownstein2007}.  The MOG theory can fit the CMB angular power spectrum and the late time matter power spectrum in cosmology~\cite{Moffat2020,Moffat2021}. In all cases no dark matter is assumed.
\begin{figure}[htpb]
  \centering
    \includegraphics[width=3.2in,height=2.0in]{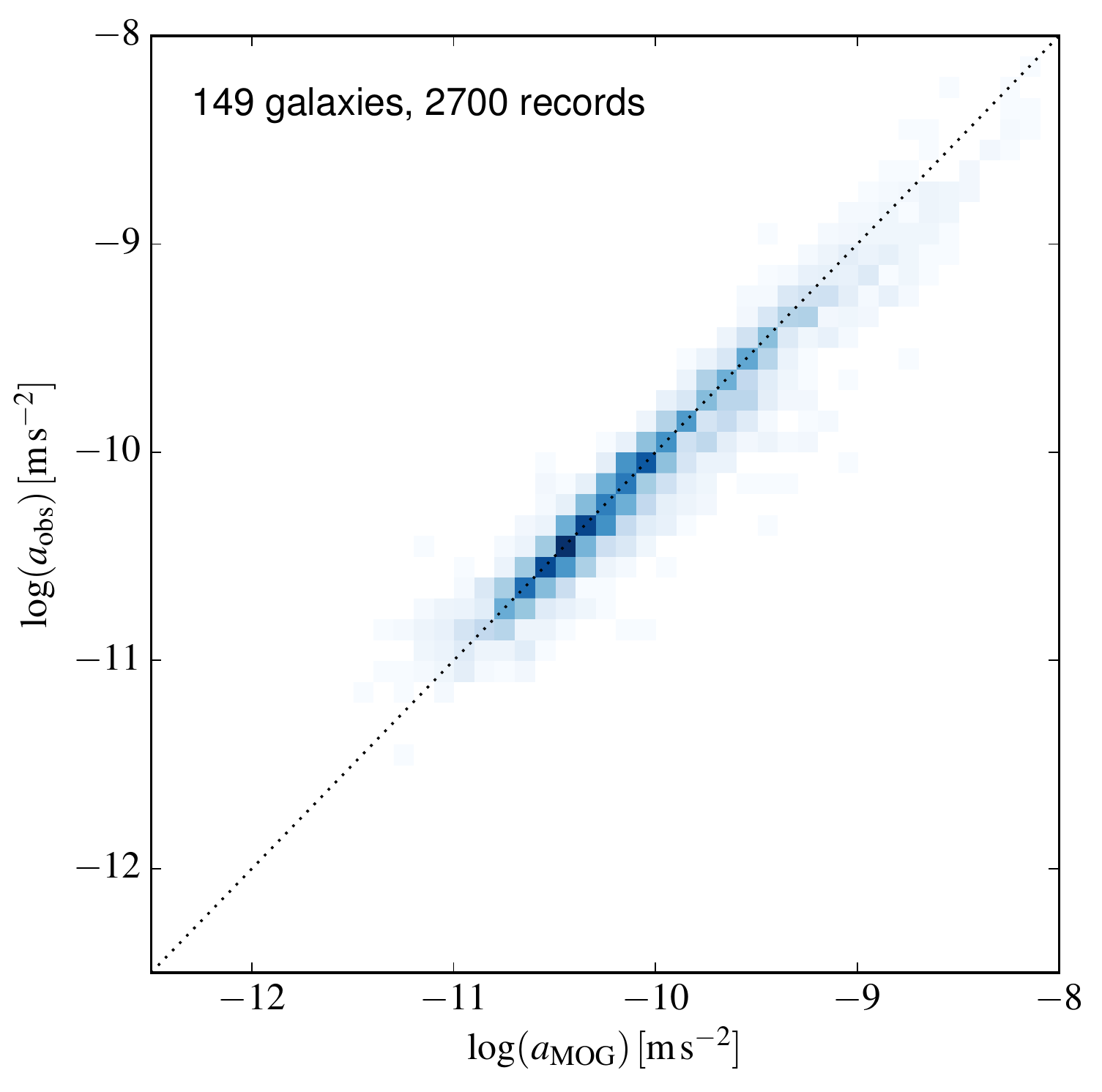}
  \caption{ MOG fit to $g(OBS)$ for 2700  data point records. }
   \label{RARMOG}
\end{figure}

Equation (\ref{MOGaccelerationlaw}) was
derived and applied to astrophysical data using the assumption that variations of $\alpha$ and $\mu$ are
ignorably small within the spacetime region being considered.  As an alternative, we consider a simplified version of MOG by instead of using Eq.~(\ref{Eq:SS}), we set $S_S=0$ and treat
$\alpha$ and $\mu$ as parameters that run, taking effectively constant values within each galaxy that depend on the scale of the system under investigation and the spatial resolution with which it is observed.  This is analogous to the running of masses and coupling parameters associated with
renormalization group (RG) flow when a condensed matter or particle physics system is observed at different scales. For $\mu$ this leads to $\mu(M)=D_0M^{-1/2}$ where $M$ the visible galactic mass and $D_0=6.25\times 10^{3}(M_{\odot})^{1/2}/kpc$; while for $\alpha$ this leads to $\alpha(M)= \alpha_{\infty}M/(M^{1/2}+E_0)^2$, where $\alpha_{\infty}=10$ and $E_0=2.5 \times 10^4M_{\odot}^{1/2}$ \cite{Moffat2009}. Comparison of the resulting MOG gravitational acceleration $a_{\rm MOG}$ with the centripetal acceleration $g(OBS)$ in Fig. [\ref{RARMOG}] shows that MOG fits  the data without dark matter~\cite{GreenMoffat2019}. For the solar system the parameter $\alpha\lesssim 10^{-9}$, so that MOG agrees with solar system experiments.

In MOG, the acceleration due to a spherically symmetric mass distribution can be calculated as~\cite{MoffatToth2019}:
\begin{align}
a_{\rm MOG}(r)=&
-\int_0^{r}dr'\dfrac{2\pi G_Nr'}{\mu r^2}\rho(r')\bigg\{2(1+\alpha)\mu r'+\alpha(1+\mu r)\left[e^{-\mu(r+r')}-e^{-\mu(r-r')}\right]\bigg\}
\nonumber\\
&-\int_{r}^\infty dr'\dfrac{2\pi G_Nr'}{\mu r^2}\rho(r')\alpha
\bigg\{\big[1+\mu r\big]e^{-\mu(r'+r)}-\big[1-\mu r\big]e^{-\mu(r'-r)}\bigg\}.
\label{4.14e}
\end{align}
We see that the integral from $r$ to $\infty$ contains external field contributions that are foreign to Newton-Einstein gravity. It can be shown that for average cosmological densities $\rho$, a computation of the acceleration $a_{\rm MOG}(r)$ yields a small additional contribution to the radial acceleration of a test particle due to an assumed homogeneous distribution of matter in space ~\cite{MoffatToth}. The external field effect modification of MOG is not an ad hoc modification of the acceleration law. However, for reasonable values of the constant density $\rho$ for baryons, the external field effect  contributions to the radial acceleration are found to be very small. Nonetheless, (\ref{4.14e}) does establish a point of principle for external field effects in MOG.

\section{Final Comments}
\label{S5}

In all three  of the alternate theories that we have discussed in this paper there are universal scales ($a_0$, $\gamma_0$, $\kappa$, $D_0$, $E_0$, and $\alpha_{\infty}$), and the successes of these theories  lie in the fact that there are also universal scales in the data themselves, with the departure of $g(OBS)$ from $g(NEW)$ universally setting in at around  $10^{-10}$ m s$^{-2}$. As well as being of order the MOND $a_0$ scale, we note that if we divide  $10^{-10}$ m s$^{-2}$ by $c^2$  we obtain $10^{-29}$ cm$^{-1}$, which is of order the conformal gravity $\gamma_0$ scale.

To gain some more insight into this scale, we note that while MOND was originally formulated so that the $a_0$ term was inertial, in \cite{Bekenstein1984} it was noted that $a_0$ could instead have been put on the gravity side. Thus we replace $\mu(x)=x/(1+x^2)^{1/2}$ and $\mu(x)=x/(1+x)$  by
\begin{eqnarray}
g(MOND)=\nu\left(\frac{g(NEW)}{a_0}\right)g(NEW),\quad \nu(y)=\left[ \frac{1}{2}+\frac{1}{2}\left(1+\frac{4}{y^2}\right)^{1/2}\right]^{1/2},\quad
\nu(y)=\frac{1}{2}+\frac{1}{2}\left(1+\frac{4}{y}\right)^{1/2}.
\label{5.1e}
\end{eqnarray}
If we describe the net effect of integrating over the $\beta^*$ and $\gamma^*$ terms in (\ref{3.6f}) as yielding a local acceleration contribution $g_{LOC}$ to the conformal gravity $g(CG)$, and describe the effect of the $\gamma_0$ and $\kappa$ terms as yielding a global acceleration contribution $g_{GLOB}$ to $g(CG)$, we can write the effect of (\ref{3.6f}) as producing an acceleration of the form 
\begin{eqnarray}
g(CG)=\nu\left(\frac{g_{LOC}}{g_{GLOB}}\right)g_{LOC},\quad \nu(y)=1+\frac{1}{y}.
\label{5.2e}
\end{eqnarray}
Similarly, we can write the MOG acceleration given in (\ref{MOGaccelerationlaw}) in the form
\begin{eqnarray}
a_{\rm MOG}(r)=\nu\left(\frac{g(NEW)}{g^{\prime}(NEW)}\right)g(NEW),\quad  \nu(y)=\frac{1}{y}-\left(\frac{1}{y}-1\right)e^{-\mu r}(1+\mu r),\quad g^{\prime}(NEW)=\frac{G}{G_N}g(NEW).
\label{5.3e}
\end{eqnarray}

As we see, in their generic forms (\ref{5.2e}) and (\ref{5.3e}) are akin to the MOND (\ref{5.1e}), and while they represent different extrapolations of Newtonian gravity, their all possessing  universal scales enables them to fit data.  However, since (\ref{5.1e}) is a phenomenological formula while (\ref{5.2e}) is derived from a fundamental theory, one can say that the derivation of the $\gamma_0$ term from cosmology in conformal gravity justifies the use of $a_0$ in MOND.

\begin{figure}[htpb]
  \centering
    \includegraphics[width=3.2in,height=2.0in]{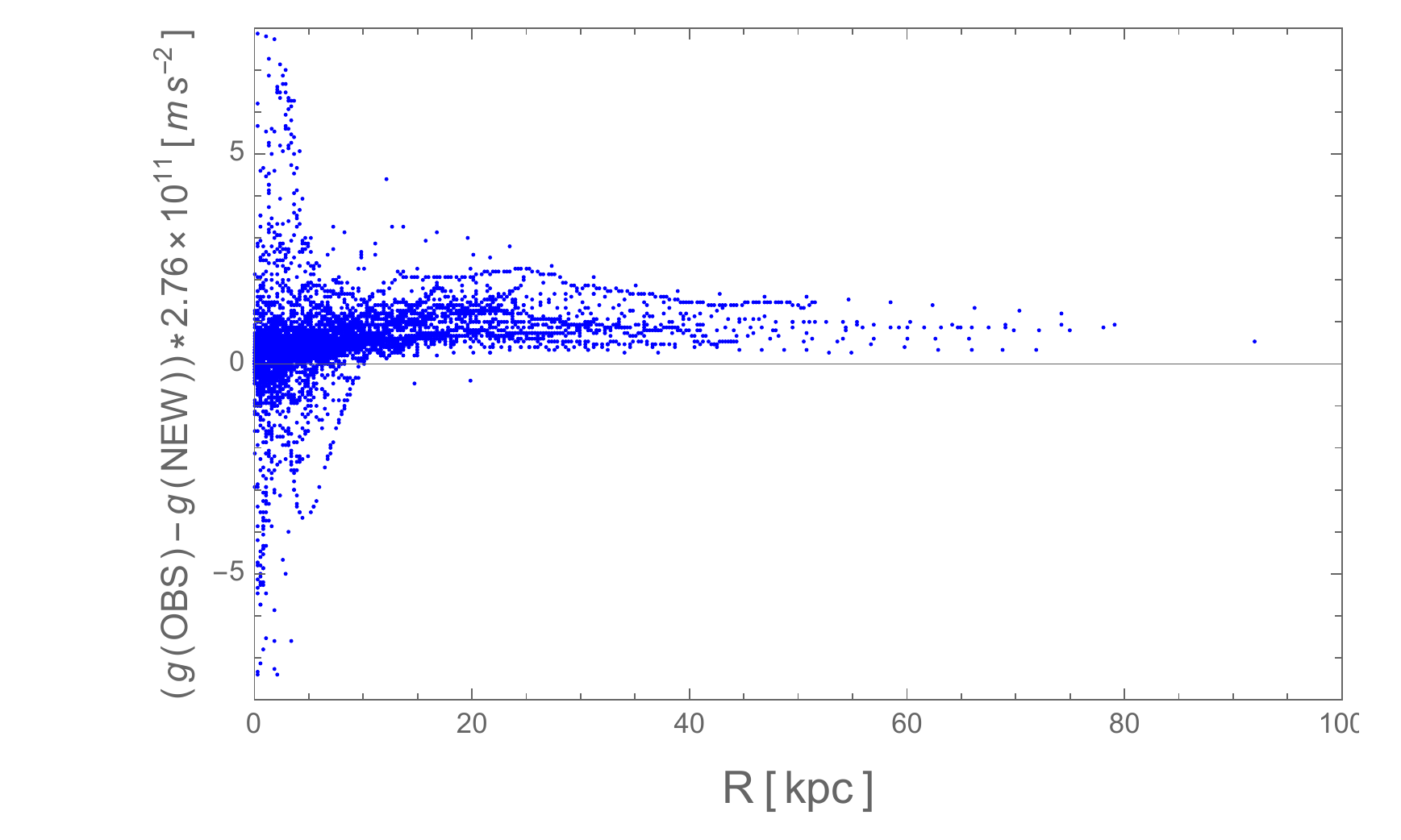}
  \caption{$g(OBS)-g(NEW)$ scaled by $\gamma_0c^2$ versus galactic radial distance $R$.}
   \label{versusr}
\end{figure}

To make the role of a universal acceleration scale manifest, instead of plotting  $g(OBS)$ versus  $g(NEW)$ we plot $g(OBS)-g(NEW)$ against the radial distance  $R$  for the 5791 data sample points, to obtain \cite{O'Brien2018}  Fig. [\ref{versusr}].
From Fig. [\ref{versusr}] we see that in every galaxy $g(OBS)$ universally exceeds $g(NEW)$ once $R$ is greater than 10 kpc, a thus natural distance scale to characterize the missing mass problem. Moreover, we see that above 10 kpc the $g(OBS)-g(NEW)$ difference is essentially a straight line with, as per (\ref{3.6f}), slope $\gamma_0c^2/2$. Since numerically $\gamma_0$ is cosmological in scale,  then, regardless of theory, the purely phenomenological  compilation of data points above 10 kpc shown in Fig. [\ref{versusr}] represents an external,  Machian-like, field effect writ large.

\begin{acknowledgments} 
The authors wish to acknowledge helpful conversations with M. A. Green, S. S. McGaugh and V. T. Toth.
\end{acknowledgments}

\end{document}